\documentclass[conference]{IEEEtran}
\IEEEoverridecommandlockouts

\usepackage[utf8x]{inputenc}
\usepackage[T1]{fontenc}
\usepackage[english]{babel}
\usepackage{microtype}
\usepackage{amsmath,amssymb,amsfonts, bm}
\usepackage{algorithmic}
\usepackage{graphicx}
\usepackage{booktabs, multirow}
\usepackage[para,online,flushleft]{threeparttable}
\usepackage[inline]{enumitem}
\usepackage{url}
\usepackage{xcolor}
\usepackage{hyperref}
\usepackage{pifont}
\usepackage{array}
\usepackage{balance}
\usepackage{setspace}

\graphicspath{{images/}}
\DeclareGraphicsExtensions{.pdf,.jpeg,.png}

\newcommand{\class}[1]{\texttt{#1}}
\newcommand{\role}[1]{\textit{#1}}

\definecolor{dark-red}{rgb}{.6, .15, .15}
\definecolor{dark-blue}{rgb}{.15, .15, .55}
\definecolor{accent1}{HTML}{24B9FC}
\definecolor{accent2}{HTML}{9ECD67}
\definecolor{accent3}{HTML}{FFA929}
\definecolor{light-bg}{HTML}{B2B2B2}

\hypersetup{
	colorlinks,
  linkcolor={dark-red},
	citecolor={dark-blue},
  urlcolor={black}
}

\newif\ifunblind
\unblindtrue

\def\BibTeX{{\rm B\kern-.05em{\sc i\kern-.025em b}\kern-.08em
    T\kern-.1667em\lower.7ex\hbox{E}\kern-.125emX}}

\begin{document}

\ifunblind

\title{Feature Maps: A Comprehensible Software Representation for Design Pattern Detection\\
\thanks{}
}
\author{\IEEEauthorblockN{Hannes Thaller, Lukas Linsbauer, and Alexander Egyed}
\IEEEauthorblockA{\textit{Institute for Software Systems Engineering}\\
\textit{Johannes Kepler University Linz, Austria}\\
\{hannes.thaller, lukas.linsbauer, alexander.egyed\}@jku.at}
}

\else

\title{Feature Maps: A Comprehensible Software Representation for Design Pattern Detection}
\author{\IEEEauthorblockN{1\textsuperscript{st} Anonymous}
\IEEEauthorblockA{\textit{Institute}\\
\textit{Affiliation, Country}\\
email@address.com}
}

\fi

\maketitle

\begin{abstract}
Design patterns are elegant and well-tested solutions to recurrent software development problems.
They are the result of software developers dealing with problems that frequently occur, solving them in the same or a slightly adapted way.
A pattern's semantics provide the intent, motivation, and applicability, describing \emph{what} it does, \emph{why} it is needed, and \emph{where} it is useful.
Consequently, design patterns encode a well of information.
Developers weave this information into their systems whenever they use design patterns to solve problems.
This work presents \emph{Feature Maps}, a flexible human- and machine-comprehensible software representation based on micro-structures.
Our algorithm, the \emph{Feature-Role Normalization}, presses the high-dimensional, inhomogeneous vector space of micro-structures into a feature map.
We apply these concepts to the problem of detecting instances of design patterns in source code.
We evaluate our methodology on four design patterns, a wide range of balanced and imbalanced labeled training data, and compare classical machine learning (Random Forests) with modern deep learning approaches (Convolutional Neural Networks).
Feature maps yield robust classifiers even under challenging settings of strongly imbalanced data distributions without sacrificing human comprehensibility.
Results suggest that feature maps are an excellent addition in the software analysis toolbox that can reveal useful information hidden in the source code.
\end{abstract}

\setstretch{0.993}
\section{Introduction}
Design Patterns (DPs) are elegant and well-tested solutions to recurrent software development problems.
\textit{Design Patterns -- Elements of Reusable Object-Oriented Software}, by Gamma et al. \cite{Gamma1995}, is the best-known collection of patterns and inspiration for many follow-ups.
They are the result of software developers dealing with problems that frequently occur, solving them in the same or a slightly adapted way.
DPs are the generalization of the different adapted implementations, such that they can be reused and applied over and over again in different situations.
They solve high-level Object-Oriented (OO) architectural problems dealing with creation, structure or behavior of a small set of classes or objects, but may also circumvent deficiencies and inflexibilities in OO languages.

Pattern descriptions are very detailed and contain their name, intent, motivation, where they are applicable, structures, participants, collaborations, and so forth~\cite{Gamma1995}.
A pattern's semantic is given by the intent, motivation, and applicability, which describe \textit{what} the pattern does, \textit{why} the pattern is needed, and \textit{where} it is useful.
Developers weave this information encoded as design patterns into their system as they use them to solve problems.
The usage of a pattern is related to specific design decisions during development.
However, fast development cycles often prohibit the documentation of these decisions and their rationales.
Similarly, the actual usage of the pattern is seldom documented.
Hence decision, rationale and their materialization in the form of the pattern's implementation are lost in the system's source code.
Retrieving this encoded information such that development, redevelopment, and maintenance can profit from it is the primary motivation of Design Pattern Detection (DPD).
For the sake of simplicity, we will use the term design pattern detection to describe the process of detecting \emph{instances} of design patterns.
DPD is especially useful for preliminary analysis in maintenance and testing scenarios where the code is unknown or undocumented.
The detected patterns hint at structures and dependencies, highlight algorithms and their moving parts, and help to find performance-critical regions.

The biggest challenge in DPD is that patterns are only a guideline for implementing a specific solution; hence each pattern can be implemented in various ways.
Each pattern implementation variant resembles the original intent of the pattern but may diverge drastically if compared to other implementation variants.
Detecting all variants and mapping the classes to pattern roles is a non-trivial task as enumerating the different variants is not sufficient for real-world setups and their inherent variations.

We use Feature Maps (FMs) as input to Random Forests (RFs) \cite{Breiman1984} and Convolutional Neural Networks (CNNs) \cite{Lecun1998a} to decide whether a given set of classes maps to the roles of a specific design pattern.
These feature maps are human-interpretable, stacked, named subtrees (micro-structures) extracted from a system's Abstract Semantics Graph (ASG) that can be used as a software representation for a small set of classes.
Results indicate that feature maps are an excellent approach to represent software, enabling robust DPD even if DP instances are highly under-represented.
More specifically, the contributions of this work are:
\begin{itemize}
	\item a new, flexible and comprehensible software representation called \emph{feature maps} that are useful for software analysis,
	\item an approach for \emph{detecting instances of design patterns in source code} by using \emph{feature maps} in conjunction with \emph{supervised machine learning},
	\item an evaluation and methodology proposal of reproducible and comparable design pattern detection.
\end{itemize}

Section~\ref{sec:background} provides background to design patterns and machine learning.
Section~\ref{sec:pipeline} describes our approach for detecting design pattern instances with feature maps.
Section~\ref{sec:evaluation} presents the experiment setup and evaluation results.
A general discussion is given in Section~\ref{sec:discussion} and threats to validity are discussed in Section~\ref{sec:validity}.
Section~\ref{sec:related work} summarizes related research and Section~\ref{sec:conclusion} concludes this work and offers prospects of design pattern detection with feature maps and machine learning.

\section{Background}\label{sec:background}
Two ingredients are crucial for the approach within this work.
First, Design Patterns (DPs) and their intricacies from a Design Pattern (Instance) Detection (DPD) perspective influence how detectors are built and evaluated.
Thus a more formal definition helps to frame the problem and its solution.
Secondly, Machine Learning (ML) and its automation capabilities concerning data modeling are essential to understand the approach and its evaluation.

\subsection{Design Patterns}\label{sec:design patterns}
\begin{figure*}[ht]
	\centering
	\includegraphics[width=\textwidth]{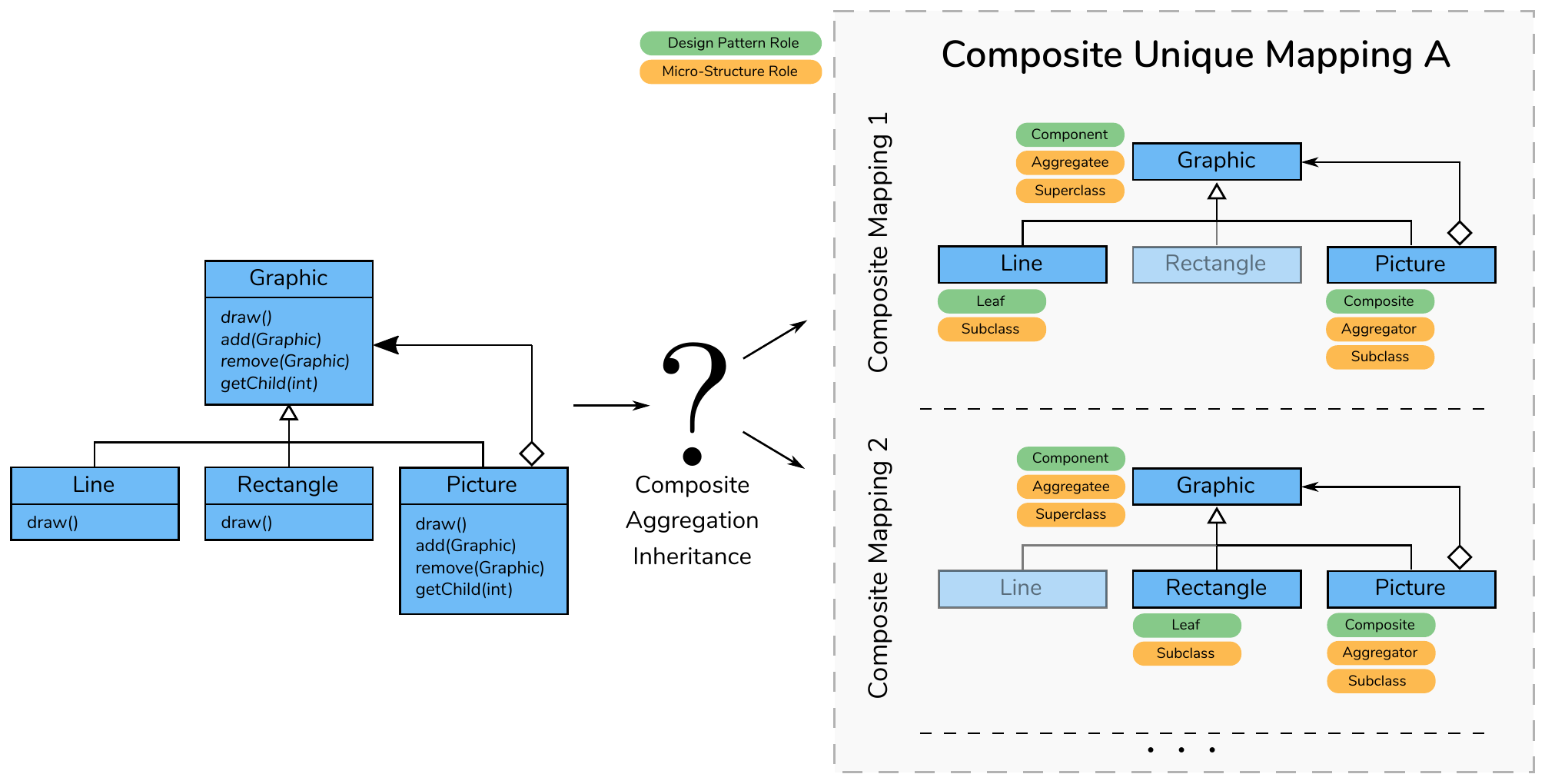}
	\caption[Basic detection process.]{
		The goal is to find a process that can reconstruct the original role mappings of a given pattern.
		Inputs are classes of a system, outputs are mappings between pattern roles and the input classes.
		A unique mapping defines a set of mappings that share the same classes mapped to the primary roles.
    Component (Graphic) is assigned to the primary role and defines the communication scheme, while \role{Composite} (Picture), \role{Leaf} (Line and Rectangle) are classes assigned to secondary roles.
		In addition to the Composite design pattern, the example also presents the detection of the Aggregation and Inheritance micro-structures.
		Similar to the detection of design patterns, micro-structures are only an assortment of classes that form a class structure.
  The Composite example is based on an example by Gamma et al. \cite{Gamma1995}.
	}
	\label{fig:introduction1}
\end{figure*}

A Design Pattern (DP) is a set of roles to which participating classes are mapped.
These classes communicate (structure and collaboration) in an organized fashion and have specific semantics (intent, motivation, and applicability) attached to them.
Design pattern (instance) detection reconstructs the original mappings between classes of a system and roles of a pattern with respect to their communication such that the attached semantics provide information about the system under inspection.

Figure \ref{fig:introduction1} (based on an example by Gamma et al. \cite{Gamma1995}) illustrates this process in which the input is a set of classes, and the output is a set of mappings between these classes and the roles of a specific pattern.
In other words, the task is to annotate a set of classes with the roles of a design pattern.
The example is a subsystem of a drawing tool that uses the Composite pattern to interact, in a uniform way, with the drawing primitives and the scene.
The Composite pattern organizes objects into a tree structure where internal nodes delegate specific calls while leaf nodes implement the actual behavior of the call.
\class{Graphic} is the superclass of \class{Picture}, \class{Line}, and \class{Rectangle} and defines a common interface for these.
\class{Picture} aggregates \class{Graphic} objects, and delegates calls to the \emph{draw}-method.
\class{Line} and \class{Rectangle} provide the actual implementation of \emph{draw}, hence are called leafs.
The goal is now to find a process that reconstructs the original mappings between classes and pattern roles indicated by the question mark in Figure~\ref{fig:introduction1}.
The left side in Figure~\ref{fig:introduction1} represents the initial state of the system before the detection process.
The right side shows the system with annotated roles after the detection process.
Mapping~1, for example, maps \class{Graphic} $\mapsto$ \role{Component}, \class{Picture} $\mapsto$ \role{Composite} and \class{Line} $\mapsto$ \role{Leaf}.

Each role mapping assigns at least one class to one role resulting in multiple possible role mappings for the same subsystem given various implementations of the different roles.
Mappings 1 and 2 differ only in one role mapping which is a common scheme in design patterns.
\emph{Primary roles} define the communication scheme within the pattern and drive the communication through a pattern's class structure, hence are often abstractions.
\emph{Secondary roles} provide the implementation for the abstractions and inherit the protocol from the primary roles, thus are commonly fluctuating in their class assignment.
A system usually provides multiple versions of the secondary roles but only a handful of different implementations for the primary roles.
All mappings that share the same primary roles belong to the same \emph{unique role mapping} representing one specific implementation of a pattern within a subsystem.
More formally, a pattern mapping is a $k$-fold relation between a set of classes $\bm{C}$ and a set of roles $\bm{R}$ with
\begin{equation}
	m_{P^k} = \{ (\bm{a}, \bm{b}) \in \bm{C}^{k} \times \bm{R}^k: \bm{a} \text{ complies with } \bm{b}\},
\end{equation}
in which $ P^k $ is a specific pattern with $ k $ roles.
Each unique role mapping reflects an equivalence class in which the \emph{primary roles} are compared.
Given a set of pattern mappings $ \bm{M}_{P^k} $ with the equivalence relation $\sim_{P^{k}}: M_{P^k} \times M_{P^k} $ in which the classes mapped to the primary roles are compared, then
\begin{equation}
	\langle m\rangle_{\sim_{P^{k}}} = \{x \in \bm{M}_{P^{k}} : x \sim_{P^k} m\}
\end{equation}
represents the $ P^k $ equivalence class from $ m $.
That is, all role mappings of pattern $P$ with $k$ roles that have the same classes $\bm{C}$ mapped to the primary roles $\bm{R}$.
In Figure \ref{fig:introduction1}, \class{Graphic} maps to the \role{Component} role representing the primary role of the Composite pattern.
\class{Line}, \class{Rectangle}, and \class{Picture} map to the secondary roles where Mappings 1-2 belong to the \emph{Unique Mapping A}.
Tertiary roles (e.g., Client) that solely function as an entry point for the pattern's communication sequence are often ignored in detection processes as they do not carry any useful information.
The concept of unique mapping carries importance in analysis settings where they explicitly state the boundaries for design pattern instances.

\subsection{Machine Learning}\label{sec:background ML}
Machine Learning (ML) describes methods that learn relationships or structure from data in an automated fashion.
Essential elements of ML are data (e.g., images or time series), the model (e.g., Random Forest~\cite{Breiman1984} or CNN~\cite{Lecun1998a}), the optimization procedure (e.g., Adam~\cite{Kingma2014}) and the evaluation procedure (e.g., Cross-Validation~\cite{Bishop2006} or test-set~\cite{Hastie2009} method).

\subsubsection{Data}
The DPD problem can be framed as a supervised classification problem in which observations are annotated with the ground truth (e.g., is an instance of Composite, is not an instance of Composite).
ML algorithms often expect the data to be independent and identically distributed (i.i.d.) meaning that the observations are mutually independent and collected in the same fashion.
Especially observation independence is vital as violating the assumption results over-optimistic model performance in classification settings.

Data can be preprocessed by standardizing or rescaling the values into a specific range.
In addition to normalizing, data permutation is often employed to increase the amount of available data or to fit models that are robust against simple transformations (i.e., reduce overfitting) \cite{Goodfellow2016b}.
For instance, rotating or mirroring an image does not change the objects it shows (e.g., rotated apple is still an apple).

\subsubsection{Model}
Many different supervised machine learning models of varying complexity and with different strengths and weaknesses exist.

\emph{Convolutional Neural Networks (CNNs)}~\cite{Lecun1998a} are fundamental building blocks in modern machine learning and are prominent for their capabilities for computer vision problems.
They learn local correlations within volumes and combine these correlations into high-level features at later stages.
For instance, nearby pixels often correlate with each other, e.g., color and texture of an apple on the top right corner of an image.
CNNs innately model such local correlations leading to good results in many domains while still having a reasonable amount of model parameters.

A \emph{Decision Tree}~\cite{Murphy2012} recursively partitions the input space via axis-parallel splits that can be represented as a tree.
Each leaf carries a response (e.g., a pattern class) while the nodes represent boolean conditions on an input dimension (e.g., is an interface $\mapsto$ \{yes, no\}).
A \emph{Random Forest (RF)}~\cite{Breiman1984} contains multiple randomly perturbed decision trees, that reduce the possible high variance of a single tree.
The trees are perturbed by fitting each on a subset of the data which creates smoother decision boundaries as multiple splits are averaged in the final prediction.

\emph{Hyper-parameter optimization} searches the space of possible parameter configurations of an ML algorithm that leads to the best performing model according to some predefined goal function.
For instance, \emph{tree depth} or \emph{partition criterion} are hyper-parameters of a decision tree.

\subsubsection{Evaluation}
It is crucial that the evaluation procedure provides a measurement of the model that is truthful, i.e., it should not over- or underestimate the performance.
Cross-Validation~\cite{Geisser1993} (CV) is a method to evaluate the generalization performance by splitting the dataset into $k$-folds.
$k-1$ folds are used to fit a model while the remaining fold is used to estimate its generalization performance.
This process is repeated $k$ times leading to $k$ models and estimates that are averaged into a global generalization performance of the ML algorithm regarding the dataset and parameters.
CV is a nearly unbiased estimator for the generalization performance except in the case of small datasets in which the evaluation variance may misestimate the generalization.
Repeated cross-validation may be used to reduce the variance, trading it for some bias.

Accuracy, Precision, Recall or the Matthews Correlation Coefficient (MCC)~\cite{Matthews1975, Powers2007} are performance metrics for classification models.
Accuracy, Recall, and Precision are frequently used metrics and need no further explanation.
MCC is a reliable and balanced performance metric for binary classification ranging from -1 to 1 describing the strength of association between model prediction and ground truth.
It provides the most accurate measurement of a (binary) model's performance even if the dataset is imbalanced (skewed distribution of prediction labels).
This does not hold for Accuracy, Precision and Recall as they do not consider all types of correctly or incorrectly predicted instances (true/false positive, true/false negatives).

\section{Design Pattern Instance Detection Pipeline}\label{sec:pipeline}
\begin{figure*}[ht]
	\centering
	\includegraphics[width=.85\textwidth]{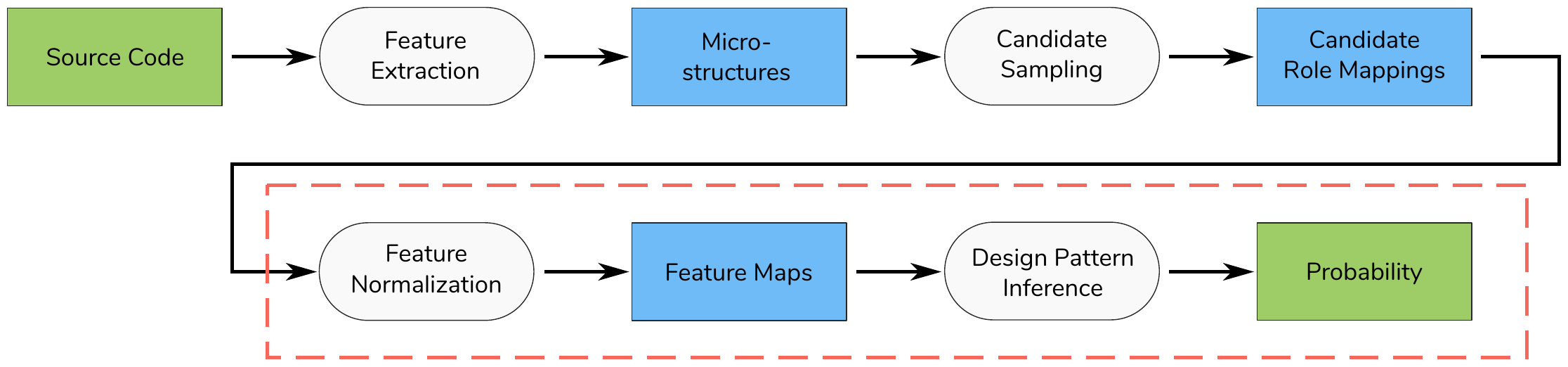}
	\caption[System overview]{
		Overview of the detection pipeline. Ellipses are processes; rectangles are artifacts.
		First, class features are extracted from the source code.
	    Then, based on the features, pattern instance candidates are sampled from the system.
	    Optionally, the features are normalized and ultimately classified by the inference method.
	}
	\label{fig:overview1}
\end{figure*}

Figure~\ref{fig:overview1} shows a typical multistage DPD pipeline similar to many DP detectors~\cite{Alhusain2013, Chihada2015, Hussain2017, Zanoni2015, Lucia2009, Tsantalis2006}.
Rectangles are artifacts while ellipses are processes.
\emph{Source Code} and \emph{Probability} are the input and output of the entire DPD pipeline while \emph{Micro-structures}, \emph{Candidate Role Mappings} and \emph{Feature Maps} are intermediate artifacts.
\emph{Feature Extraction} extracts distinctive features from the source code.
\emph{Candidate Sampling} reduces the total number of mappings that need to be evaluated.
\emph{Feature Normalization} transforms the features into a homogeneous space such that the \emph{Design Pattern Inference} can reason efficiently about it.

\subsection{Feature Extraction}\label{sec:feature extraction}
Feature extraction deals with the inherent complexity of programming languages by extracting high-level concepts that later stages can use to successfully find pattern instances (see Figure~\ref{fig:overview1}).
DPs describe classes and their loose arrangements and communication paths.
Consequently, extracted features ideally capture these arrangements and their relationships to improve reasoning in later stages.

We use Micro-Structures (MSs)~\cite{ArcelliFontana2011} as features, which are very small DPs (usually one or two roles).
A MS describes a general structural or behavioral property within or between a set of classes (e.g., self-reference, inheritance or method calls).
Casting MSs into an easily readable form allows developers and algorithms to effectively comprehend these properties.
This includes fast discovery of inheritance and call dependencies but also complex aspects like the implementation of a Composite or Decorator pattern.

The most outstanding difference between MSs and DPs is that the former can be detected using logic or pattern matching, i.e., their size prohibits variance in their actual implementation.
MS detectors are sub-graph filters retaining only sub-graphs that match their specified predicate.
For example, $Inheritance(sup, sub): \mathbb{T}^2,  (sup, sub) \mapsto Ancestor(sup, sub)$, describes the logic to filter for the \emph{Inheritance} MS.
Given two Type nodes, $sup$ has to be an ancestor of $sub$ in order to fulfill the $Inheritance$ predicate.
The result of such an MS detector are sub-graphs from the program's ASG that are annotated with the MS roles illustrated in Figure \ref{fig:introduction1} for the \emph{Inheritance} and \emph{Aggregation} MS.
\class{Graphic} $\mapsto$ \role{Superclass} and \class{Picture} $\mapsto$ \role{Subclass} for one possible \emph{Inheritance} MS instance.
A complete catalog of Micro-Structures and their detectors is given in our previous work~\cite{Thaller2016}, which is a refined and extended catalog based on Maggioni's work~\cite{Maggioni2009}.
The catalog is made up of three sub-catalogs: Elemental Design Patterns~\cite{Smith2011}, Design Pattern Clues~\cite{ArcelliFontana2011b} and Micro-Patterns~\cite{Gil2005}.
Each sub-catalog was defined independently with different motivations and goals, but all of them prove valuable in the process of DPD as Arcelli Fontana et al. \cite{ArcelliFontana2011a} concluded in a series of experiments.

\subsection{Candidate Sampling}\label{sec:candidate sampling}
Candidate Sampling uses the extracted features (and sometimes the \emph{Source Code}) to find potential candidates of design pattern instances in a program's ASG.
As discussed, role mappings link concrete classes to specific pattern roles which span a search space of $\binom{n}{k}$ potential mappings where $n$ is the number of available classes and $k$ the number of roles that need to be mapped for a pattern.
Hence, finding potential candidates is vital since a full-search is impossible for non-trivial systems.
We used Heuristic Search~\cite{Thaller2016} that we proposed in our previous work as candidate sampler.
Heuristic Search checks for graph isomorphism~\cite{Wilson1986} between a very general description of the pattern and classes in the program under inspection.
It uses primary roles as an entry point to iteratively search for secondary roles that fit the pattern description.
Heuristic Search reduces the usual $2^{O(\sqrt{n \log(n)})}$  isomorphism search~\cite{Babai1983} to a search that is linear with the number of types.
For Composite, heuristic search executes the following steps:
\begin{enumerate*}
  \item Collect all types $\bm{sup}$ that are super-types as \role{Component} ($\bm{sup} \mapsto$ \role{Component});
  \item Collect all types $\bm{sub}$ that are sub-types from $\bm{sup}$ and aggregate $\bm{sub}$ as \role{Composite} ($\bm{sub} \mapsto$ \role{Composite});
  \item Collect all types $\bm{sib}$ that are sub-types from $\bm{sup}$ and do not aggregate $\bm{sup}$ as \role{Leaf} ($\bm{sib} \mapsto$ \role{Leaf}).
\end{enumerate*}
Each found role mapping is a candidate that is later checked in the design pattern inference stage for their correct consistency with the pattern.
A drawback to this heuristic search is that it might miss potential pattern instances.
The advantage is that these are simple and efficient algorithms to find potential candidates.

\subsection{Feature Normalization}\label{sec:frn}
Feature-Role Normalization (FRN) is the main contribution of this work.
It is used to normalize micro-structures into a fixed sized matrix called Feature Map (FM).
FRN allows the usage of a wide variety of existing statistical methods while still retaining the interpretability for manual analysis via software engineers.
Figure \ref{fig:overview1} shows the pipeline in which first the features (micro-structures) are extracted followed by the sampling of candidates from the system.
A candidate mapping is a set of classes that map to the design pattern roles which may, may not or may partially map to several micro-structures.
Each of the potentially mapped micro-structures may have a different number of roles, resulting in an inhomogeneous feature space.
This inhomogeneous feature space is problematic in the inference step as most machine learning methods cannot handle irregularly structured data.
Feature-Role Normalization presses this inhomogeneous feature space into a homogeneous two-dimensional feature space that is comprehensible and maintains enough information to be useful for engineers and learning algorithms.
More specific, the goal of the normalization, is to take $ n=|\bm{f}| $ features each having $ m_{fi} = |\bm{r}_{fi}| $ roles and map them to $ k = |\bm{r}_{di}| $ design pattern roles such that the result is a fixed-sized matrix.
FRN provides one approach to this problem by creating an $ n \times k $ matrix where rows represent micro-structures and columns design pattern roles.
Values in the feature map, in the context of DPD, are globally unique role identifiers ($ id \in \mathbb{N} $), where no two roles of any pattern have the same values associated.
That said, roles might be shared across multiple features if they share similar semantic information.
Zeros represent missing features meaning that the class mapped to the role does not participate in the micro-structure.

\begin{figure}[t]
	\centering
	\includegraphics[width=.9\columnwidth]{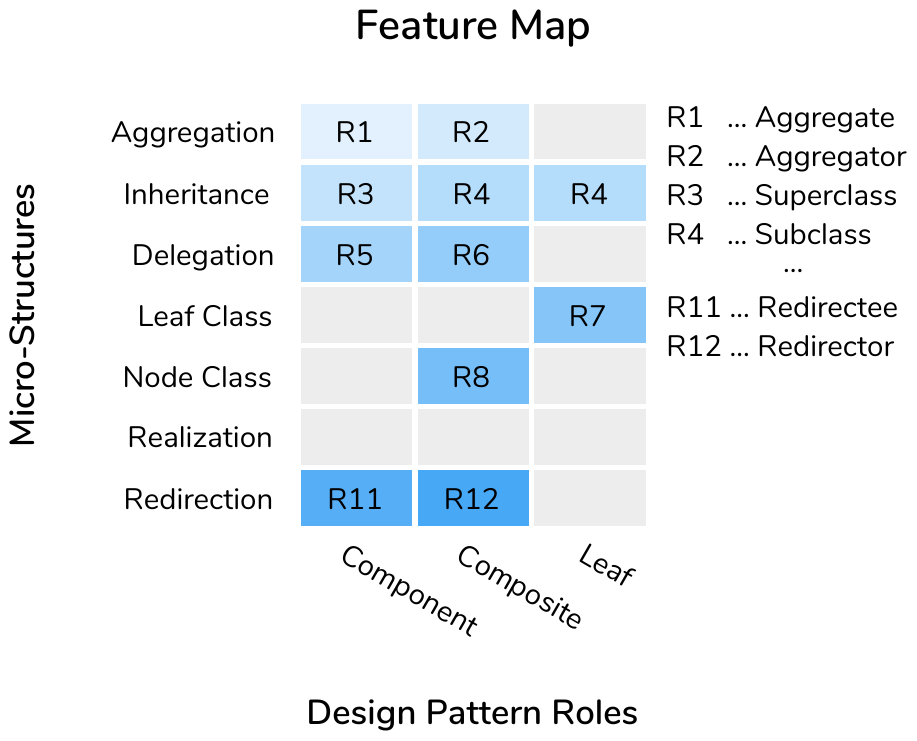}
	\caption[Feature Map]{
  A (condensed) feature map is a stack of micro-structures for a set of classes (represented by their Decorator roles in this case).
  Values represent unique role identifiers with $id \in \mathbb{N}$ or any other property related to micro-structures (e.g., number of occurrences of the micro-structure).
  The coloring in this example is a gradient based on the role identifiers. Gray spots represent absent values (zeros).
	}
	\label{fig:feature_map}
\end{figure}

Figure~\ref{fig:feature_map} illustrates an example of a feature map describing classes that make up a Composite pattern.
Values are role identifiers colored by a simple gradient, and empty gray spots are absent mappings (zeros).
Rows depict an illustrative subset of the micro-structures (features) and columns the design pattern roles to which a set of classes are mapped.
The class assigned to \role{Component} is a \role{Superclass} with its children classes assigned to \role{Composite} and \role{Leaf}.
\role{Composite} aggregates a \role{Component}, delegating and redirecting method calls to it.

This short example mirrors how software engineers would, in a slow fashion, go through the source code of a project to find possible instances of a Composite pattern.
Feature maps simplify this process by compacting the relationships between multiple classes into new high-level concepts (e.g., a design pattern) and offer a straightforward way of visualizing them.
Naturally, the visualization can be extended by recoloring, grouping or filtering the entries of a feature map, or by visualizing another source code property than identifiers (e.g., number of invocations between roles or complexity measures).
This makes FMs a strong visualization technique for software analysis.
They provide a compact view of specific aspects of a small set of classes allowing a quick overview of their relationships and class characteristics.
Furthermore, the matrix structure has clear advantages regarding automation as many methods, e.g., machine learning algorithms, are designed to work with flat data.

Detection processes that operate on structured data may circumvent the use of FMs for DPD by directly operating on the ASG.
Examples are Support Vector Machines~\cite{Burges1998} (with tree-kernels), Recursive Neural Networks~\cite{Gori2003}, sequence-to-sequence learning models like Long Short-Term Memory Networks~\cite{Hochreiter1997} or Graph Convolutional Neural Networks~\cite{Kipf2016}.
These learning algorithms and their respective models might provide possible performance advantages in the context of DPD by naturally handling the ASG and the MSs.
However, some lack computational efficiency while others lack interpretability of the classification decisions.

\subsubsection{Information Preservation Issues in Feature Maps}\label{sec:issues}
FRN normalizes the inhomogeneous feature space caused by micro-structures and their roles into a fixed-sized feature map.
The resulting feature map is a sparse representation of attributes and relationships within and between nodes in an ASG.
While inspired by adjacency matrices, feature maps are not a proper and full representation of the subgraphs; hence they suffer from two specific issues: noise and collisions.

\begin{figure*}[ht]
	\centering
	\includegraphics[width=\linewidth]{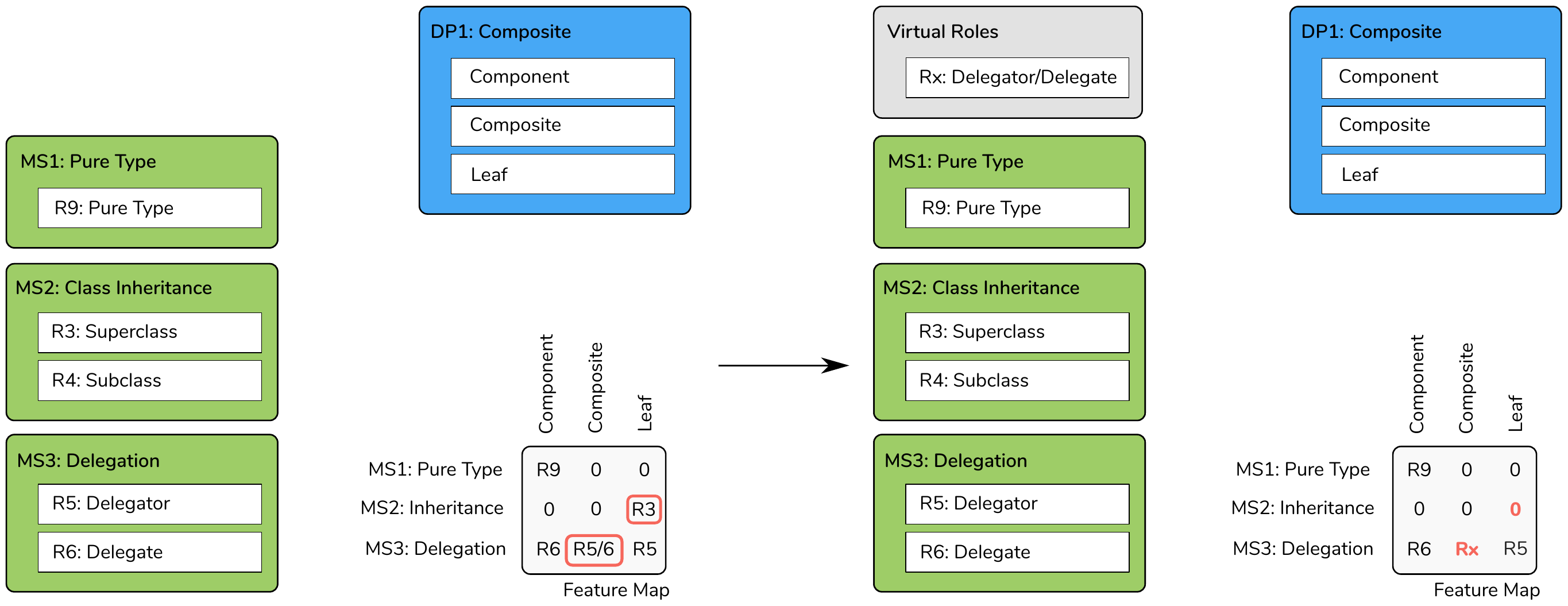}
	\caption[Feature-Role Normalization Issues]{
		The left side illustrates an example of noise (R3) and role collisions (R5/6) within feature maps.
		Noise is introduced by features that relate to classes that do not participate in the actual mapping.
		Role collisions occur if a feature is present multiple times with the same set of mapped classes.
		The right side provides a solution to both issues where mappings to classes that do not participate in the actual mapping are ignored (forced to 0),
		and role collisions are solved by introducing virtual roles (Rx) that represent multiple roles at the same time.
	}
	\label{fig:frn_issue}
\end{figure*}

A feature introduces \emph{noise} if it has at least two roles and only one of its roles map to a class within the FM.
Figure \ref{fig:frn_issue} illustrates this problem with the Composite pattern and the features Pure Type, Inheritance and Delegation.
On the left side, the class assigned to \role{Leaf} is a \role{Superclass}.
However, neither of the other Composite roles participate in this inheritance hierarchy (both are 0).
The \role{Superclass} role of the \role{Leaf} class exists in the project but does not carry any meaningful information in the context of the current FM mapping.
These out of context values inflate the feature map with irrelevant information and introduce noise reducing the usability of FMs.
A simple solution to the problem is to force entries to zero in cases where features depict a relationship, but none of the other mapped classes participate in it.
Figure \ref{fig:frn_issue} depicts this solution on the right side where \emph{R3} becomes $0$.

A feature suffers from a role \emph{collision} if it has at least two roles and multiple instances of the feature trace to the same classes.
Figure \ref{fig:frn_issue} (left) illustrate a collision where two instances of the Delegation MS are present in the classes assigned to \role{Component}, \role{Composite}, and \role{Leaf}.
\role{Composite} delegates to \role{Component} and \role{Leaf} delegates to \role{Composite}.
The multi-assignment for \role{Composite} is the result of the class-level abstraction and compactness of FMs.
One possible solution is to introduce virtual roles that describe two or more roles like \role{Rx} in Figure \ref{fig:frn_issue} on the right.
Virtual roles incur a certain amount of information loss (direction of relationship) especially if multiple virtual roles have to be used for a feature in a FM.
However, virtual roles also allow to express self-loops, for instance if a class aggregates itself (it is \role{Aggregator} and \role{Aggregate} at the same time).
These self-loops are helpful in many cases, for example, $50\%$ of the Singleton pattern instances contain \role{Creation Site / Type To Create} indicating that the class is the factory for itself.
This information is essential in the context of Singleton detection.

\subsection{Design Pattern Inference}\label{sec:dpi}
Design Pattern Inference is the last step of the DPD pipeline (Figure~\ref{fig:overview1}) and receives feature maps as input.
During training time the models and their parameters are optimized toward a labeled dataset of feature maps.
During production time the models only return the probability that an FM belongs to the model's optimized pattern or not (no labels, no model parameter modifications).
The detailed intricacies of the training and production time are subject to the selected model class and may differ quite drastically.
For instance, a CNN is optimized via a gradient-based algorithm while decision trees are constructed by splitting the input dimension values via an information theoretical criterion (training time).

\section{Design Pattern Instance Detection Study}\label{sec:evaluation}
This study frames the process of DPD (see Figure~\ref{fig:introduction1}) as a binary classification problem.
Each design pattern is detected by a separate model that provides the probability that the input FM belongs to the design pattern.
Naturally, multi-label classification, i.e., multiple patterns per observation (FM), is supported by feeding the same FM into different models.

Given this setup, we conducted several experiments with FMs in the context of design pattern instance detection.
For better reproducibility and comparability we evaluated only the last two stages of the DPD pipeline highlighted in Figure \ref{fig:overview1}.
Experience shows that reimplementing the entire pipeline of a DPD approach including the intermediates (e.g., sampler results) while making the detection results comparable is nearly impossible.
That said, the entire pipeline from Figure \ref{fig:overview1} was implemented, and details about the specific implementation of micro-structures and candidate samplers are given in our previous work~\cite{Thaller2016}.
The following evaluation and experimental methodology should be seen as a minimal effort to guarantee comparable research in the future.

\subsection{Controlled Variables}\label{sec:controlled variables}
Detecting design pattern instances can be a daunting task as there are numerous ways to encode software (e.g., feature maps), select potential candidates (candidate sampling), and build classification models (e.g., RFs or CNNs).
In total, the study controls for 7 Experiment Parameters (ExP).

\begin{enumerate}
  \item\label{param:pattern} \textbf{Pattern:} Four widely used patterns are considered \{\textit{Singleton, Template Method, Composite, Decorator}\}.
  This selection is based on previous studies~\cite{Tsantalis2006, Uchiyama2011, ArcelliFontana2011a, Zanoni2015} but also represents an even selection regarding their categorical distribution (creational, structural and behavioral patterns).

  \item\label{param:role count} \textbf{Role Count:} A confounding variable is the number of roles of the pattern.
  The expectation is that patterns with one role are easier to detect than patterns with 2 or 3 roles.
  However, more roles also imply more conditions through which better precision in the classification can be achieved.
  The selected patterns have a linear role count distribution with \{\textit{1, 2, 3, 4}\} roles for \{\textit{Singleton, Template Method, Composite, Decorator}\} excluding tertiary roles.

  \item\label{param:model} \textbf{Classification Model:} \{\textit{Random Forest, CNN}\} were selected as classification model classes.
  The rationale for selecting CNNs is their natural ability to handle volumes, i.e., to process FMs while leveraging their structural information.
  RFs were selected as they are a popular and efficient choice in many ML tasks and are more light-weight regarding hyper-parameters in contrast to CNNs.

  \item\label{param:npcr} \textbf{Negative-Positive Candidate Ratio:} Non-trivial software systems expose a combinatorial number of possible class mappings that can function as pattern candidates.
  This means a practically infinite number of pattern counterexamples can be found within a system.
  The candidate samplers (Section \ref{sec:candidate sampling}) that mitigate this issue favor \textit{recall} leading to imbalanced distributions of positive and negative examples.
  Negative-Positive Candidate Ratio (NPCR) captures this skew in the data.
  The study explores NPCRs of \{\textit{1, 2, 4, 6, 8, 10}\}.

  \item\label{param:data augmentation} \textbf{Data Augmentation:} Shuffling of rows was used as data augmentation method to increase the dataset size and classification difficulty.
  This inhibits the classifier from learning specific relationships between rows, instead of their content.
  A permutation count of \{\textit{0, 1, 5, 10}\} states how often the dataset was copied, rows augmented and added to the dataset.
  Permuting rows \emph{and} columns was not considered as it would result in a nearly random matrix destroying any structural information captured in feature maps.

  \item\label{param:budget} \textbf{Optimization Budget:} A budget of \{\textit{200}\} evaluations was available for each model class (RF or CNN).
  These evaluations were used to tune their hyper-parameters (such as the number of units within layers, or depth of tree).
  The optimization selected hyper-parameter configurations that maximized the MCC.

  \item\label{param:independence} \textbf{Instance Independence:} Describes the confounding variable that instances within a unique mapping correlate thus result in too optimistic classification results if not properly handled.
  Using standard cross-validation mixes mappings from the same unique mapping into training and test-folds making the folds not independent.
  This leads to too optimistic evaluation results as the training examples \textit{leak} information between training and test phase.
  We used \{\textit{project-fold cross validation}\} to avoid information leaks (correlations between samples) during the trials.
\end{enumerate}

\subsection{Response Variables}\label{sec:response variables}
The result of each experiment trial was the generalization performance of the classifier evaluated through cross-validation and measured by \{\textit{Accuracy, Precision, Recall, F1, Matthews Correlation Coefficient}\}.
MCC was selected as the primary evaluation metric as it provides the most unbiased single number metric for binary classification.

\subsection{Data Source}
A total of $4$ widely used design patterns were selected for the study, each having peer-reviewed classification examples.
These peer-reviewed classifications are given by nine projects that are part of the Pattern-like Micro-Architecture Repository (P-MARt\footnote{\url{http://www.ptidej.net/tools/designpatterns/index_html\#2}} 04/10/19)~\cite{Gueheneuc2007}.
Applications within the repository range from modeling and drawing tools to static analysis and refactoring frameworks, therefore its diversity is a good representation for the real world.
Table \ref{tab:pmart_projects} contains the distribution of design pattern instances across projects.
\textit{Original} are the mapping counts in the exploded form provided by the dataset.
\textit{Revised} are the mappings after a manual cleanup of the data.
Some design pattern instances trace to classes not contained in the available source code repositories (e.g., java.awt, javax.swing, or third-party libraries) and were removed.
Projects that did not include the target pattern were also not considered during the experiments (for the respective pattern) as controlling for imbalance (NPCR) is impossible, e.g., JRefactory for Template Method.
At last, the Netbeans project was excluded as we were not able to parse it.
Fixing the source code pieces in question was deemed futile as they were too numerous introducing potential bias in the evaluation.
\textit{Unique} are the unique mappings after the revision.

\begin{table}
	\centering
	\caption[P-MARt Projects Overview]{
		The 9 projects contained in the P-MARt dataset.
		For every project the original and the revised (excluding instances outside of the available source code) number of design pattern instances (i.e. mappings) as well as the number of unique mappings is given.
	}
	\label{tab:pmart_projects}
	\begin{threeparttable}
		\resizebox{\columnwidth}{!}{
			\begin{tabular}{@{}l  r r r r@{}}
				\toprule[.1em]
				\multirow{2}{*}{\textbf{Project}} & \multicolumn{4}{c}{\textbf{Original / Revised / Unique}} \\
				 & Singleton & Template Method & Composite & Decorator \\
				\cmidrule[.075em]{1-5}
				JHotDraw 		& 2 / 2 / 2		  & 68 / 34 / 2		& 3840 / 960 / 1	& 176 / 44 / 1\\
				JRefactory 	& 2 / 2 / 2 		& 0 / 0 / 0			& 0 / 0 / 0				& 0 / 0 / 0 \\
				JUnit 			& 2 / 2 / 2 		& 0 / 0 / 0			& 222 / 74 / 1	  & 396 / 132 / 1 \\
				Lexi 				& 2 / 2 / 2 		& 0 / 0 / 0			& 0 / 0 / 0				& 0 / 0 / 0 \\
				MapperXML 	& 3 / 3 / 3			& 88 / 44 /4		& 340 / 85 / 1		& 0 / 0 / 0 \\
				Netbeans 		& - / - / - 		& - / - / -			& - / - / -				& - / - / - \\
				Nutch 			& 1 / 1 / 1			& 14 / 7 / 2		& 0 / 0 / 0 			& 0 / 0 / 0  \\
				PMD 				& 0 / 0 / 0 		& 1 / 0 / 0  		& 3 / 0 / 0  			& 0 / 0 / 0  \\
				QuickUML 		& 1 / 1 / 1 		& 0 / 0 / 0  		& 120 / 30 / 2 		& 0 / 0 / 0  \\
				\cmidrule[.075em]{1-5}
				\textbf{Total} & \textbf{13 / 13 / 13} & \textbf{171 / 85 / 8} & \textbf{4525 / 1149 / 5} & \textbf{572 / 176 / 2} \\
				\hline
				\bottomrule[.1em]
			\end{tabular}
		}
	\end{threeparttable}
\end{table}

\subsection{Procedures}

First, $67$ micro-structures were extracted (Section~\ref{sec:feature extraction}) and possible candidates sampled (Section~\ref{sec:candidate sampling}) as shown in Figure~\ref{fig:overview1}.
The micro-structure extraction and candidate sampling were based on our previous work~\cite{Thaller2017}.
Feature-Role Normalization (Section~\ref{sec:frn}) created the feature maps for each candidate and pattern.
Cells within the FM were globally unique role identifiers ($id \in \mathbb{N}$) which received no further preprocessing such as rescaling or standardization.
Further, no specific assignment strategy was employed for identifiers, i.e., role identifiers were assigned in their encountered order resulting in a total value range of $[0, 161]$ including the virtual roles.

The controlled variables from Section~\ref{sec:controlled variables} were executed in a systematic fashion resulting  in the response variables from Section~\ref{sec:response variables}.
All features (rows) were included.
First, the FMs were collected into datasets, and negative-subsampling was applied according to the NPCR ratio (ExP~\ref{param:npcr}).
The dataset was then copied, permutated, and added $k$ times (ExP~\ref{param:data augmentation}).
FMs were provided to CNNs as-is and linearized (row-wise) into a vector for RF.
Then the models were fitted (ExP~\ref{param:model}) and evaluated using \textit{project-fold cross-validation} (ExP~\ref{param:independence}).
This procedure was repeated $200$ times in search of the optimal hyper-parameters (ExP~\ref{param:budget}) and for each pattern (ExP~\ref{param:pattern}).

\subsection{Experiment Results}\label{sec:experiment_results}
\begin{table*}
	\centering
	\caption[Detection performance]{
	Average test-performance of the Convolutional Neural Network (CNN) and Random Forest (RF) based detections.
	}
	\label{tab:designPatternDetection8}
	\begin{threeparttable}
		\resizebox{.7\linewidth}{!}{
		\begin{tabular}{@{}l l r r r r r @{}}\toprule[.1em]
			\textbf{Model} & \textbf{Pattern} & \textbf{Accuracy}  	& \textbf{Precision} & \textbf{Recall} 	& \textbf{F1} 	& \textbf{MCC}\\
			\cmidrule[.075em]{1-7}
			\multirow{4}{*}{{CNN}} 	& Singleton 				& $0.88 \pm 0.09$ & $0.82 \pm 0.12$ & $0.73 \pm 0.12$ & $0.76 \pm 0.10$ & $0.72  \pm 0.14$\\
											& Template Method     		& $0.82 \pm 0.09$ & $0.63 \pm 0.18$ & $0.62 \pm 0.20$ & $0.58 \pm 0.14$ & $0.51  \pm 0.12$\\
											& Composite 				& $0.94 \pm 0.02$ & $0.86 \pm 0.11$ & $0.80 \pm 0.14$ & $0.82 \pm 0.10$ & $0.79  \pm 0.10$\\
											& Decorator 				& $0.81 \pm 0.10$ & $0.69 \pm 0.15$ & $0.80 \pm 0.17$ & $0.72 \pm 0.13$ & $0.60  \pm 0.16$\\
			\cmidrule[.075em]{1-7}
				\multirow{4}{*}{{RF}}& Singleton 					 & $0.81 \pm 0.07$ & $0.63 \pm 0.12$ & $0.59 \pm 0.12$ & $0.59 \pm 0.06$ & $0.47  \pm 0.06$\\
											& Template Method 		 		 & $0.85 \pm 0.04$ & $0.66 \pm 0.21$ & $0.58 \pm 0.16$ & $0.59 \pm 0.15$ & $0.49  \pm 0.12$\\
											& Composite 					 & $0.93 \pm 0.04$ & $0.89 \pm 0.10$ & $0.71 \pm 0.21$ & $0.77 \pm 0.16$ & $0.79  \pm 0.14$\\
											& Decorator 					 & $0.45 \pm 0.06$ & $0.10 \pm 0.15$ & $0.09 \pm 0.13$ & $0.09 \pm 0.12$ & $-0.35 \pm  0.10$\\
			\hline
			\bottomrule[.1em]
		\end{tabular}
		}
	\end{threeparttable}
\end{table*}

\begin{figure*}[ht]
	\centering
  	\includegraphics[width=.83\textwidth]{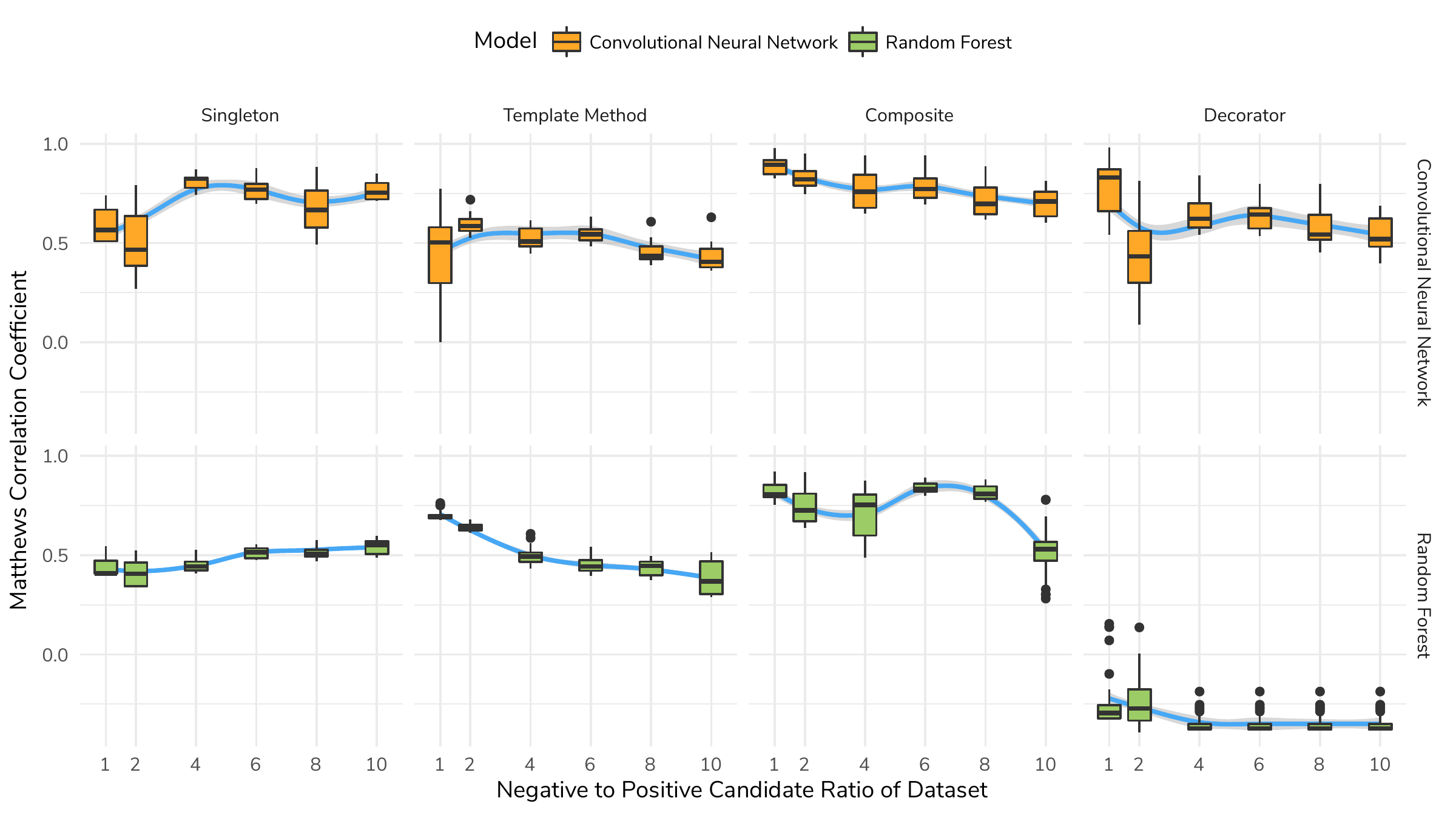}
	\caption{
    Each boxplot represents the top-30 fitted models for a given pattern and model, illustrating the trend across imbalanced datasets (NPCRs).
    Trends indicate that Feature Maps are a robust representation for DPD even in highly imbalanced settings.
	}
	\label{fig:mcc evaluation}
\end{figure*}

The experiment parameters led to a total of $384$ trials in which $76 800$ classifiers were fitted.
The bulk of these classifiers were part of the optimization budget (ExP~\ref{param:budget}).
In total, the best performing top-30 classifiers (including ties) for each Pattern, Model, and NPCR combination was selected for the evaluation.
For instance, the boxplot in Figure \ref{fig:mcc evaluation} for the Singleton pattern at an NPCR of 1 represents the top-30 classifiers of this category.
Furthermore, Table \ref{tab:designPatternDetection8} contains the average test performance (marginalized over NPCR) for each pattern and model.

Using the Evans guidelines~\cite{Evans1996} for interpreting correlations, we see that CNN models have on average a \emph{strong} performance with a Median (Med) $0.646$ and an Interquartile Range (IQR) of $0.528$ to $0.772$.
The worst median performance is given by the Template Method classifiers reaching a \emph{moderate} $mcc_{Med} = 0.51;~mcc_{IQR} =(0.43, 057)$).
Best median performance is given by the Composite pattern with \emph{strong} $mcc_{Med} = 0.79;~mcc_{IQR} =(0.71, 0.85)$.
The total variance in these estimates is acceptably low and would cause in the worst case a performance drop of $mcc_\Delta  =0.16$ degrading it to a \emph{moderate} classifier.
CNNs are rather robust against data imbalance with an average NPCR 1 to 10 change of $mcc_\Delta = -0.064$.

RF models are quite similar to CNN models for patterns with low role count but degrade quite drastically for patterns with four roles.
RFs have on average a \emph{moderate} performance with $mcc_{Med}=0.48;~mcc_{IQR}=(0.29, 0.64)$.
At least a \emph{moderate} ($mcc_{Med}=0.47;mcc_{IQR}=(0.41, 0.52)$) up to a \emph{strong} performance ($mcc_{Med}=0.79;mcc_{IQR}=(0.67, 0.83)$) is given, excluding the Decorator models that were systematically worse ($mcc_{Med}=-0.35;mcc_{IQR}=(-0.38, -0.29)$) than a random model ($mcc = 0$).
Again, the best performance as reached for Composite patterns with $mcc_{Med}=0.79;mcc_{IQR}=(0.67, 0.83)$.
On average, RF models exposed a variance of $mcc_\Delta = 0.14$ (Composite).

Figure \ref{fig:mcc evaluation} shows the effect of NPCRs on the classifier's performance.
Both, CNN and RF suffer from imbalance except in the case of the Singleton pattern.
CNNs are on average (all patterns) more robust against imbalance with an average NPCR 1 to NPCR 10 difference of $mcc_{\Delta}=0.064$.
However, they degrade for Composite and Decorator.
The average difference for RFs is $mcc_{\Delta} = -0.16$.

Tests for independence~\cite{Kruskal1952} regarding MCC between CNN models with different permutation counts (ExP~\ref{param:data augmentation}) were close to significant $\chi (3, 730)=7.5245, p < 0.057$.
Permutation passes for RF models were significant with $\chi (3, 745)=75.117, p < 3.42^{-16}$.
There was a significant effect concerning the NPCR for both model types with CNN ($\chi (5, N = 730) = 43.553, p < 2.85e^{-8}$) and RF ($\chi (5, N = 745) = 18.041, p < 0.002$).

\section{Discussion}\label{sec:discussion}
The presented results in Section~\ref{sec:experiment_results} fit with the intuition that
\begin{enumerate*}
  \item patterns with more roles are easier to detect,
  \item models have a decline in performance with larger NPCR,
  \item FMs fit well in the framework of CNNs.
\end{enumerate*}

Patterns with more roles inherently have more conditions that have to be met as each role describes some specific behavior a participating class must fulfill.
This avoids \textit{unintentional} implementations of these patterns and their detection because of a more elaborate structure and communication flow.
Template Method, for example, describes a very general and freely used concept of abstract classes that defer implementation details of an algorithm to subclasses.
Not only is this implementation technique used without actually having the intention to implement the Template Method pattern, but also pre-labeling of the dataset in this regard might prove challenging because of its general applicability.
In contrast, Composite or Decorator use inheritance in conjunction with redirection techniques forcing the classifier to focus not only on inheritance but also the redirection aspect between multiple classes.
Both patterns achieved excellent results that are quite robust across different NPCRs using CNNs (low degradation and low variance).
However, RF, while similar to CNN for Singleton and Template Method, strongly degrades with patterns that have more roles (both variance and bias).
This is to be expected as FMs encode relationship information in a 2-dimensional format (matrix) which needs to be serialized (loss of structural information) into a vector.
Furthermore, serialization with patterns that have, e.g., four roles (Decorator), result in a $67 \times 4 = 268$ dimensional input vector.
We did not explore the invertibility of the negative correlation (systematic mistakes) that RFs made for the Decorator.
However, a simple inversion of the predictions might not generalize as expected and is open for further research.
Nevertheless, RFs are still a good and fast solution for patterns with 1-3 three roles as the performance in Figure \ref{fig:mcc evaluation} shows.

The degrading performance of models with higher imbalance is often an issue.
The minority class (class with lesser samples) is outweighed and too seldom encountered during training of the classifier in order to learn the essential aspects of it.
Figure~\ref{fig:mcc evaluation} shows that the performance degrades with higher NPCR.
Still, the underlying trends are rather shallow except for Template Method.
The increase in performance for the Singleton pattern is most likely related to the amount of available data.
With only $13$ examples it is expected that the classifiers use the extra negative samples to make the overall predictions more robust.
In total, the degradation of performance over NPCRs is less extreme than initially anticipated.
Though, it would need an additional study to attribute this effect properly.

FMs work well in the framework of CNNs because of their natural processing of volumes.
CNNs are able to partially replicate the human DPD process via feature maps described in Section \ref{sec:frn}.
For example, a $3 \times 3$ filter can combine Aggregation, Inheritance, and Delegation across Component, Composite and Leaf in the first convolutional step from Figure \ref{fig:feature_map}.
The result would be a higher level internal feature (of unknown quantity and semantics) representing this combination.
Each layer then computes an even more abstract interpretation of the original feature leading to a good prediction.
The same does not necessarily hold for RFs as the structural information (relationships between DP roles) would need to be recovered from the serialized vectors while still focusing on the content.
We think that this natural advantage of CNNs resulted in the overall better performance especially for patterns with more roles.

Permutation passes increased the performance for RF classifiers and acted as additional regularization for CNNs without any performance loss (nor significant gain).

An advantage of NPCR (or generally imbalance) analysis is that it allows for direct comparison with other results via linear interpolation between NPCRs (as long as the datasets are the same).
Zanoni et al.~\cite{Zanoni2015} conducted a similar study (same dataset) where they used multitudes of different classification models to find DPs.
They report Accuracy for their \emph{best} performing models without explicitly reporting the $NPCR$ values.
We sketch a comparison by interpolating our results to match their $NPCR$ (computed form the ZeroR model) and compare our \emph{average} model accuracies with their \emph{best} model accuracies.
Their best result for the \textit{Singleton} pattern ($NPCR \approx 1.66$) was given by an RF model with $acc=0.93$.
Our RF is $20\%$ worse with $acc=0.73$, and the CNN is $19\%$ worse with $acc=0.77$.
Their best \emph{Composite} ($NPCR = 3$) model ($\nu-SVC RBF$ \cite{Smola2004}) reaches $acc=0.81$ where both of our models are better (CNN $acc=0.93$ and RF $acc=0.90$).
Their best \emph{Decorator} ($NPCR \approx 1.48$) model given by an RF with $acc=0.82$ which is quite similar to our average CNNs with $acc=0.79$.
No data for \emph{Template Method} is available.
These results show how well FMs are suited for DPD given that we compared the average performance and not the best performing models.
Besides, the authors used ordinary 10-fold cross-validation in which instances from the same unique mapping may leak information between evaluation and training folds.

\section{Threats to Validity}\label{sec:validity}
An internal threat to validity is given by the dataset (P-MARt 04/10/19~\cite{Gueheneuc2007}) we used to train and evaluate our approach.
We inspected many projects and their respective labeled design patterns in order to understand the misclassification during the experiments.
Many of these projects are already outdated using old Java versions that forced the developers to take strange roundabouts in their implementations.
Consequently, it is possible that some design patterns would be implemented differently today.
For instance, the Singleton pattern is in modern systems most likely implemented via dependency injection frameworks.
Furthermore, we could not understand some of the labeled pattern instances as they seem to diverge too far from the original design pattern definition.

The size of the dataset poses a threat, especially for the Singleton pattern as the model may overfit.
To mitigate this thread we used repeated cross-validation to estimate the generalization performance on unseen data, added data augmentation (increases amount of observations), and included additional model specific regularization methods.
Random forests naturally regularize with an increased number of trees.
We used Dropout~\cite{Srivastava2014}, kernel and activation regularization (L1, L2~\cite{Murphy2012, Goodfellow2016b}), and early stopping~\cite{Prechelt1998a} to regularize the CNNs.

Our experiment design tried to eliminate most external threats by using multiple patterns with different numbers of roles, with multiple bootstrapped datasets and different NPCRs.
However, it only evaluates the performance of one pattern with $k$ number of roles, and generalizing it to all patterns with $k$ roles may only be possible on a limited scale.
Nevertheless, the broad sample of models still provides valuable insight into the methodologies performance in many different settings.

\section{Related Work}\label{sec:related work}
The DPD research community has a long and active history with various approaches, tools, and methodologies.
Several interests can be distilled from DPD and that are reflected in Figure \ref{fig:overview1}.
A coarse classification of these interests is feature extraction, intermediate representations, candidate sampling, and inference methods.
This work focused on intermediate representations and inference methods.

Zanoni~\cite{Zanoni2015}, with whom we compared our results, uses micro-structures, clustering, and various machine learning approaches to find design patterns.
Along with Arcelli, they proposed a framework that uses these techniques called MARPLE~\cite{ArcelliFontana2011}.
Most influential work for feature maps and our approach is given by Tsantalis et al. \cite{Tsantalis2006}, in which they extract adjacency matrices from the ASG reflecting a specific aspect.
An example of an aspect would be Generalization or Association, and in such a way they may be seen as micro-structures.
The approach itself compared the adjacency matrices from patterns with the matrices extracted from the system via an iterative similarity scoring algorithm proposed by Blondel et al. \cite{Blondel2004}.
Their approach provides nearly always a recall and precision of 1 for the patterns detected in this work.
However, their evaluation is only based on three manually inspected projects (JHotDraw, JRefactory, and JUnit).
The authors accounted for subjective bias through manual inspection but did not mitigate its possible impact.
Finally, they report that their methodology suffers from computational inefficiency caused by the (adjacency) matrices and the matching algorithm.

One big issue in the DPD community is that results are hard to reproduce and compare because of the multistage nature of detectors.
Fontana et al.~\cite{ArcelliFontana2012} give an attempt to make results comparable.
Nevertheless, the approach is a community-driven web-based benchmarking system that suffers, like many other community-based approaches, from the cold start phenomenon, i.e., it provides not enough upfront benefit to justify its usage in the first place.
To improve this situation, we employed an evaluation strategy that decouples the last stage of the DPD pipeline simplifying the reproducibility and comparability of the results to the availability of the dataset that candidate samplers produce.
Furthermore, we provided a comprehensive approach to evaluate detectors in a more general fashion than peak performance models.

\section{Conclusion}\label{sec:conclusion}
This work presented Feature Maps (FMs), how they are computed via Feature-Role Normalization (FRN) and used for design pattern detection.
Feature maps themselves can be understood as flexible and comprehensible source code representation useful beyond DPD.
For DPD, FMs provide a representation that allows for robust detection performance even if the datasets are strongly imbalanced.
In conclusion, FMs do not only help to extract the information developers weave into their systems, but also provide means to represent and comprehend them in a compact fashion.

In the future, we are planning to extend the study on FMs in the context of DPD by applying the methodology to more design patterns and a bigger dataset.
Futhermore, we are planning to compare the methodology to learning algorithms that handle graph representations natively.

\section*{Acknowledgments}
The research reported in this paper has been supported by the Austrian Ministry for Transport, Innovation and Technology, the Federal Ministry of Science, Research and Economy, and the Province of Upper Austria in the frame of the COMET center SCCH.

\balance
\clearpage
\setstretch{1}
\bibliographystyle{IEEEtran}
\bibliography{references}

\end{document}